\documentclass[twoside,slac_one]{revtex4}
\usepackage{graphicx}
\usepackage{fancyhdr}
\usepackage{amsmath} % American Mathematics Society standards
\usepackage{bm}% bold math
\usepackage{amsxtra}
\usepackage{amssymb}
\usepackage{amsthm}
\usepackage{latexsym}
\usepackage{lscape}
\usepackage[stable]{footmisc}

\begin{document}

\title{Measurements of Spin Correlation in $t\bar{t}$ Events at D0}


\author{Kenneth Bloom (for the D0 Collaboration)}
\affiliation{Department of Physics and Astronomy, University of
  Nebraska-Lincoln, Lincoln, NE, USA}

\begin{abstract}
  Two recent measurements by the D0 Collaboration of spin correlation in
  $t\bar{t}$ production using 5.4~fb$^{-1}$ of Tevatron $p\bar{p}$ collider
  data are presented.  Both rely on the dilepton final state of
  $t\bar{t}$.  One measurement relies on full reconstruction of the top
  quark kinematics, and the other makes use of leading-order matrix
  elements to characterize the kinematics.  The latter measurement is the
  first ever to have sufficient analyzing power to exclude the
  no-correlation hypothesis.
\end{abstract}

\maketitle

\thispagestyle{fancy}


%%%%%%%%%%%%%%%%%%%%%%%%%%%%%%%%%%
\section{Introduction}
In the $p\bar{p} \to q\bar{q}$ process, the quarks that are produced are
unpolarized, but their spins are correlated.  This is required by angular
momentum conservation in the strong interaction.  In general, this
correlation is unobservable, as the hadronization process involves the
emission of gluons that can flip the spins of the quarks.  But the top
quark provides a laboratory for studying the correlation.  The short
lifetime of top, about $5 \times 10^{-25}$~s, is shorter than the timescale
for strong processes, so top decays before fragmentation and spin flips can
occur.  Thus, the original spin orientation is preserved, and is passed to
the decay products.  It should then be observable through a study of the
kinematics of the decay products.

A measurement of top-quark spin correlation is a test of top-quark
properties and also a probe of new physics.  The very observation of the
correlation could in principle be used to set an upper limit on the top
lifetime.  Should top have a non-standard decay (such as $t \to H^+ b$), or
non-standard production mechanism (through decays of stop pairs, or a $Z'$
resonance), a non-standard correlation would be observed.  The correlation
is ultimately a subtle effect -- the theme of this presentation -- but
there is now enough Tevatron data to explore it.

\section{About correlation}
At the Tevatron, the primary production mode of $t\bar{t}$ is through
$q\bar q \to t\bar t$ with an $s$-channel gluon.  (This is in contrast to
the LHC, where the initial state is primarily $gg$).  The $q$ and $\bar q$
must have opposite helicity to couple to that gluon, and that forces the
$t$ and $\bar t$ to have their spins pointing along the beamline.  A
correlation strength can be defined based on the number of $t\bar{t}$ pairs
with their spins pointing in the same direction,
\begin{equation}
A = \frac
{N_{\uparrow\uparrow} + N_{\downarrow\downarrow} -
 N_{\uparrow\downarrow} - N_{\downarrow\uparrow}}
{N_{\uparrow\uparrow} + N_{\downarrow\downarrow} +
 N_{\uparrow\downarrow} + N_{\downarrow\uparrow}}.
\end{equation}

But the spin orientation must be defined with respect to a quantization
axis.  In the measurements described here, the beamline axis, defined as
the direction of the colliding hadrons in the zero-momentum frame of the
$t\bar{t}$ system, is used.  This choice is intuitive, easy to construct,
and optimal for $t\bar{t}$ produced at threshold.  With this choice of
quantization axis, next-to-leading order QCD calculations predict $A =
0.777^{+0.027}_{-0.042}$~\cite{bib:Avalue}.

\begin{figure*}[ht]
\centering
\includegraphics[width=135mm]{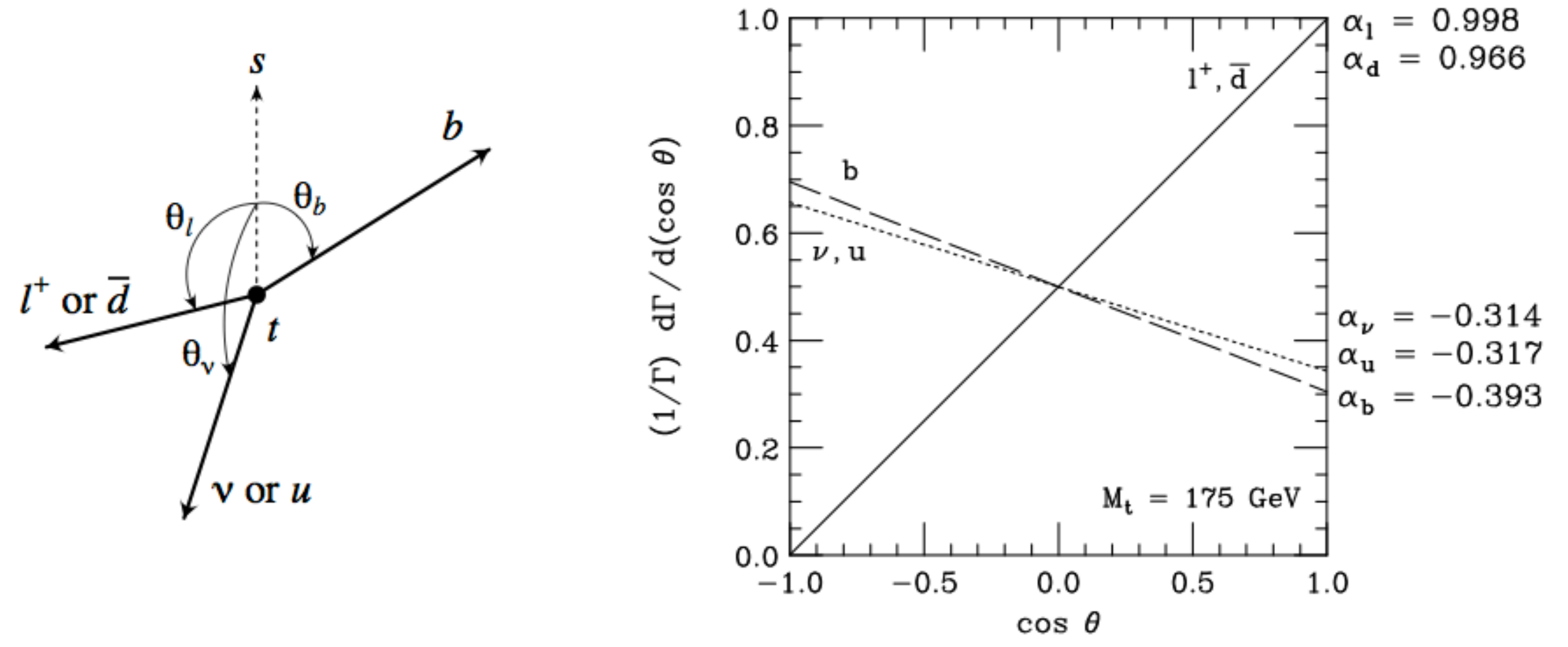}
\caption{Illustration of the top decay angles (left) and dependence of the
  decay rate on the angles (right)~\cite{bib:mahlon}.} \label{fig:angles}
\end{figure*}

The spin orientation of the top is then passed to its decay products.  The
differential angular decay distributions of the top daughters is given by
\begin{equation}
\frac{1}{\Gamma}\frac{d\Gamma}{d\cos \theta_i} = \frac{1}{2}(1 + \alpha_i
\cos\theta_i),
\end{equation}
where $\cos\theta_i$ is the angle between the $i$th top daughter and the
spin of the top.  Different decay products have different correlation
strengths, as indicated by $\alpha_i$.  This is illustrated in
Figure~\ref{fig:angles}.  In the case of a leptonic $W$ decay, the lepton
has the greatest analyzing power, and for a hadronic $W$ decay it is the
down-type quark\footnote{In the original observation of parity violation,
  it was the angular distribution of electrons from nuclear beta decay that
  were observed.  What would the history of our understanding of the weak
  interaction have been had the correlation with the nuclear spin direction
  not been so strong?}.  In both cases, $\alpha \simeq 1$.

Thus, the doubly-differential cross section as a function of the decay
angles of decay products from two different quarks in $t\bar{t}$ is given
by
\begin{equation}
\frac{1}{\sigma}\frac{d^2\sigma}{d\cos\theta_1 d\cos\theta_2} =
\frac{1}{4} (1 - A\alpha_1\alpha_2 \cos\theta_1 \cos\theta_2).
\end{equation}
To study the spin correlation, one looks for correlation between the
directions of decay products from the two different top decays.  In the
case of both tops decaying leptonically, $\alpha_1\alpha_1 \simeq 1$, and
we write $A\alpha_1\alpha_2 \equiv C \simeq A$.

\section{Experimental situation\protect\footnote{This presentation was given at the
    very end of a long day of top-physics talks, and by unanimous consent
    of the audience most of the material in this section was not presented, as it
    was deemed redunant.}}
The measurements described below were performed at the Fermilab Tevatron, a
$p\bar{p}$ collider operating at $\sqrt{s}=1.96$~TeV.  Run~II of the
collider has been in progress since 2001, with almost 12~fb$^{-1}$ of
integrated luminosity delivered.  The spin-correlation measurements make
use of 5.4~fb$^{-1}$.  The data was recorded by the D0 detector, which
consists of silicon and fiber trackers inside a 2~T solenoid, a liquid
argon-uranium calorimeter, and muon trackers and scintillators inside
toroids.

At the Tevatron, 85\% of $t\bar{t}$ production arises from $q\bar{q}$
annihilation and the remaining 15\% from $gg$ interactions.  Each top quark
decays to $Wb$ nearly 100\% of the time, and the final states are
characterized by the $W$ decay modes.  The three final states are
all-hadronic, lepton plus jets and dilepton.  In the order listed, the
final states have decreasing rate and increasing number of neutrinos and
signal purity.

The dilepton state in particular is characterized by two high-$p_T$
leptons, missing momentum due to the escaping neutrinos, two hadronic jets
from $b$ decays, and perhaps additional jets due to initial- and
final-state radiation.  For the purpose of the spin-correlation
measurements, the dilepton state has the best analyzing power and most
accurate measurement of the decay-product (lepton) directions, but the
worst statistical power, compared to other final states that could be
considered.

\section{Event selection}
The two measurements described have the same event selection.  A
$t\bar{t}$-enriched sample is chosen by selecting events with two
high-$p_T$, isolated, opposite-charge leptons.  Only electrons and muons
are considered as leptons, so the dilepton pairs can be either $ee$, $e\mu$
or $\mu\mu$..  There must also be at least two high-$p_T$ jets.  To
suppress backgrounds, a large scalar sum of the lepton and jet $p_T$ values
is required in the $e\mu$ channel, and significant missing energy in the
$ee$ and $\mu\mu$ channels.  Backgrounds from $Z/\gamma^*$ (diboson) events
are modeled by leading-order Monte Carlo samples, and normalized to
next-to-next-to-leading (next-to-leading) order cross sections.
Instrumental backgrounds arise from misidentified $\pi^0$ and $\eta$ decays
in electron samples and real muons in jets that appear to be isolated in
muon samples; both of these are modeled with complementary data samples.
The selected data sample is about 70\% pure in $t\bar{t}$ events, as shown
in Table~\ref{tab:events}.

\begin{table}[ht]
\begin{center}
\caption{Estimates of contributions of various physics processes to the
  selected dilepton sample.}
\begin{tabular}{cccccc}\hline
$t\bar{t}$ & $Z/\gamma^*$ & Diboson & Instrumental & Total & Observed\\
$341 \pm 30$ & $93 \pm 15$ & $19 \pm 3$ & $28 \pm 5$ & $481 \pm 39$ & 485\\
\hline
\end{tabular}
\label{tab:events}
\end{center}
\end{table}

\section{Analysis I: Template-based}
D0 has performed two different measurements of the spin correlation with
this event sample.  The first is template-based, in which the decay angles
are calculated in each event, and the distribution of angles is modeled
by a sum of templates representing the distributions for correlated and
uncorrelated spins~\cite{bib:meas1}.  This technique has been used
before~\cite{bib:old1}, but with a much smaller data sample.

To observe the correlation, one must measure the angle between the lepton
and the beamline (which is used as the top spin quantization axis) in the
zero-momentum frame of the $t\bar{t}$ system, which requires a full
reconstruction of the decay.  A total of eighteen quantities are needed to
specify the final-state configuration, but because of the two undetected
neutrinos only twelve are measured.  Constraining the decay kinematics to
the values of the top and $W$ masses provides four additional pieces of
information, but that still leaves two missing.

The ``neutrino weighting'' technique is used to solve the remaining
kinematics.  Two values of the neutrino $\eta$ (where $\eta =
-\ln\tan(\theta/2)$ and $\theta$ is the polar angle measured from the
beamline) are randomly sampled from the neutrino $\eta$ distribution as
predicted from $t\bar{t}$ Monte Carlo simulations.  These values are then
used to solve for the implied $t\bar{t}$ kinematics.  This allows a
determination of the product of decay angles $\cos\theta_1\cos\theta_2$ and
of the neutrino momenta.  The $\cos\theta_1\cos\theta_2$ value is then
weighted by the consistency of the determined neutrino momenta with the
measured missing transverse energy in the event.  The sampling is repeated
many times, and weighted mean of all solutions obtained is then used as the
estimator of $\cos\theta_1\cos\theta_2$.

\begin{figure*}[ht]
\centering
\includegraphics[width=80mm]{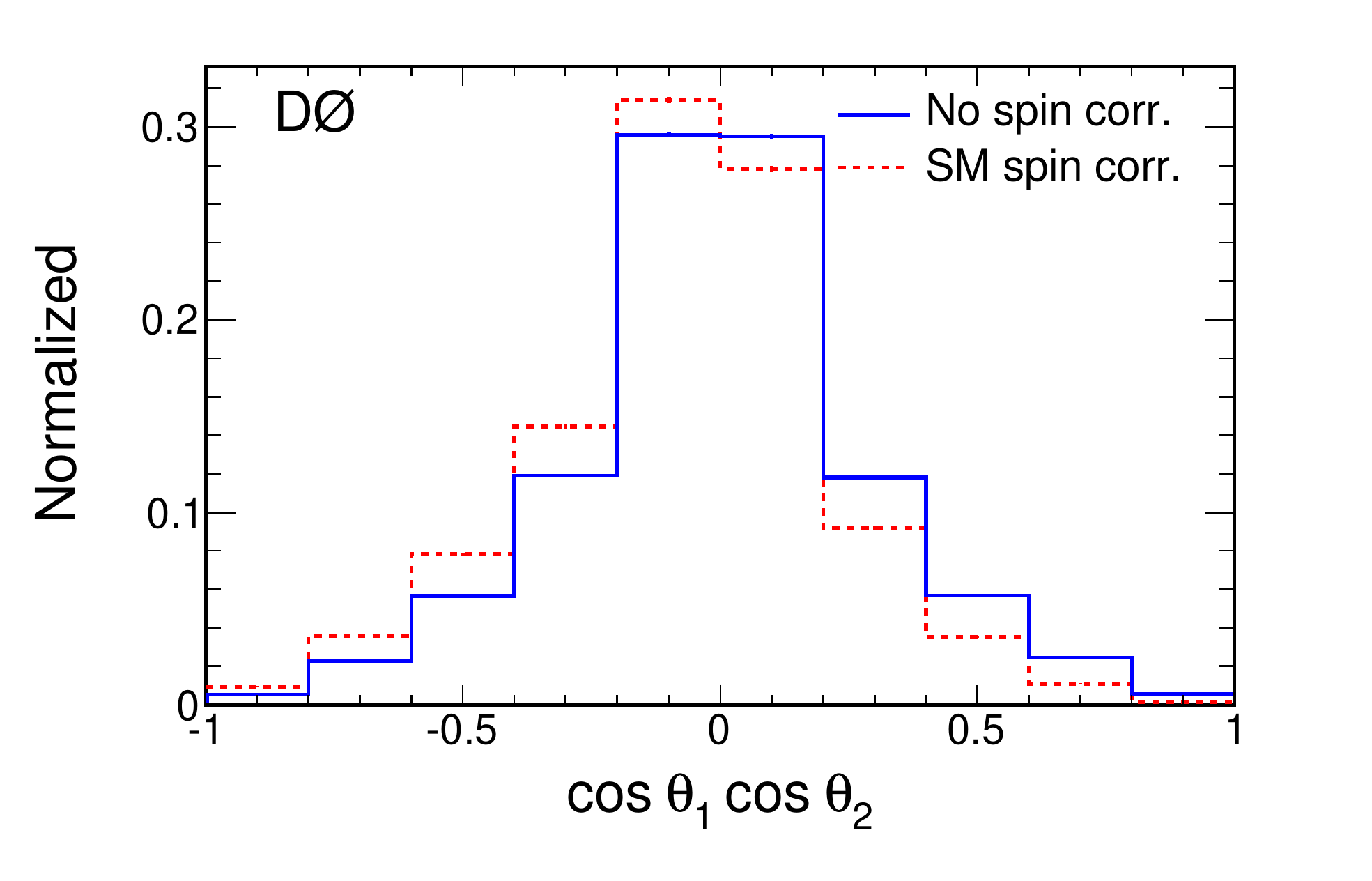}
\includegraphics[width=80mm]{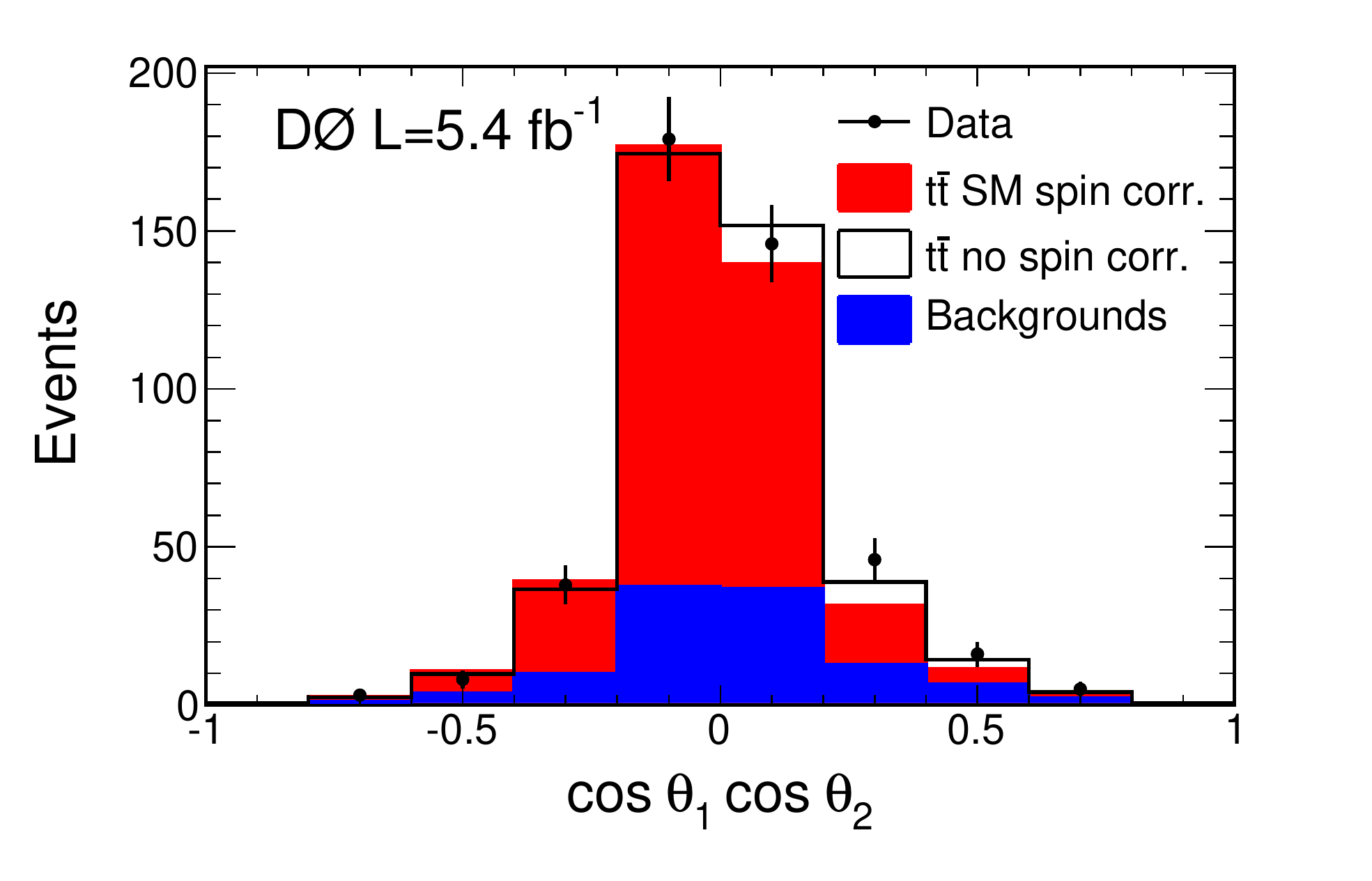}
\caption{Left: The distribution in $\cos\theta_1\cos\theta_2$ for a  sample
  including the NLO QCD spin correlation (C = 0.78) (red dashed line) and
  with no spin correlation (C = 0) (blue solid line) at the parton level,
  generated using MC@NLO.  Right: The distribution in $\cos\theta_1\cos\theta_2$ for the entire dilepton event sample. The summed  signal, including NLO QCD spin correlation (C = 0.78) (red) and all backgrounds (blue) are compared to data. The open histogram is the  prediction without spin correlation (C = 0).} \label{fig:template}
\end{figure*}

$t\bar{t}$ events are simulated using the MC@NLO
generator~\cite{bib:MCNLO}, in which the spin correlation can be turned on
or off straightforwardly.  Then, with the appropriate weighting of simulated
samples, $\cos\theta_1\cos\theta_2$ distributions for any value of $C$ can
be generated.  The left panel of Figure~\ref{fig:template} shows the
expected distribution of $\cos\theta_1 \cos\theta_2$ at parton level for
$t\bar{t}$ with no spin correlation ($C = 0$) and standard-model (SM) spin
correlation ($C = 0.78$).  The distribution is symmetric when there is no
correlation, and shifted slightly towards negative values of $\cos\theta_1
\cos\theta_2$ when there is correlation.

The right panel of Figure~\ref{fig:template} shows the distribution
observed in the data, along with the appropriately-normalized background
distribution and the distributions expected for the cases of no
correlation and SM correlation.  On the face of it, it seems
hard to distinguish the two cases.  To make a quantitative statement, the
most likely value of $C$ is obtained with a binned maximum-likelihood fit.
Systematic uncertainties are incorporated to the fit as nuisance
parameters, and the overall $t\bar{t}$ cross section is treated as a free
parameter to avoid biases.

\begin{figure*}[ht]
\centering
\includegraphics[width=80mm]{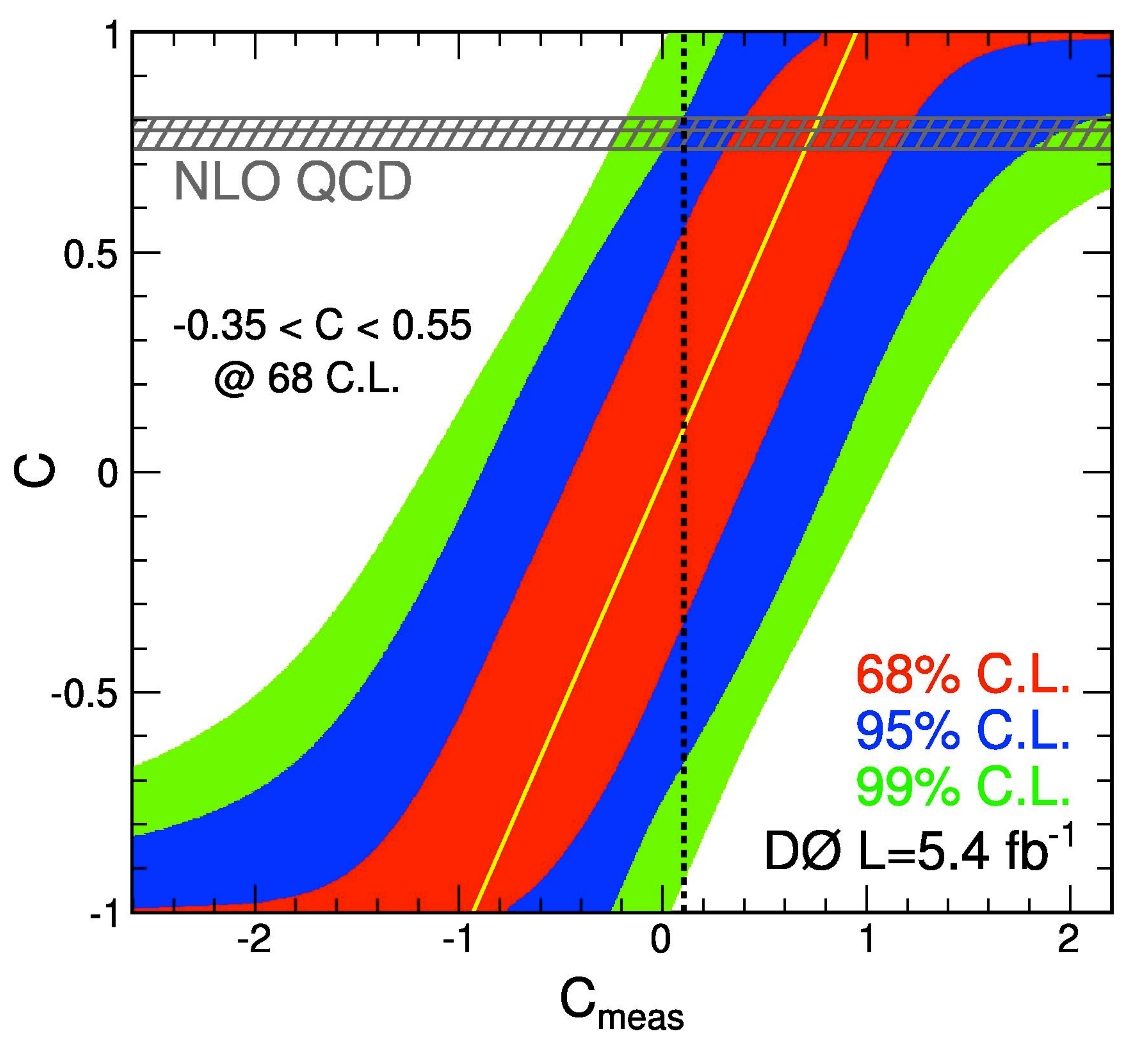}
\caption{The 68\% (inner), 95\% (middle), and 99\% (outer)
C.L. bands of $C$ as a function of $C_{\mathrm{meas}}$ from likelihood
fits to MC events for all channels combined.  The yellow line
indicates the most probable value of $C$ as a function of
$C_{\mathrm{meas}}$, and represents the calibration of the method.
The vertical dashed black line depicts the measured value
$C_{\mathrm{meas}}=0.10$. The horizontal band indicates the NLO QCD
prediction of $C=0.777^{+0.027}_{-0.042}$.}
\label{fig:templateFC}
\end{figure*}

A Feldman-Cousins-based frequentist approach~\cite{bib:FC} is used to set
confidence limits on $C$ as a function of the measured value.  These are
shown in Figure~\ref{fig:templateFC}.  The measured value is
$C_{\mathrm{meas}} = 0.10 \pm 0.45$, which can be compared with the
expected value of $C = 0.78$.  We find $-0.66 < C < 0.81$ at 95\%
confidence level.  Uncertainties on the central value are shown in
Table~\ref{tab:template_errors}.  Statistical uncertainties dominate by
far; the leading systematic uncertainty arises from the limited Monte Carlo
statistics in generating the templates.  This can obviously be remedied
when statistical uncertainties are reduced with more data.

\begin{table}[ht]
\begin{center}
\caption{Summary of uncertainties on $C_{\mathrm{meas}}$.}
\begin{tabular}{ccc} \hline \hline
 Source  &        $+$SD    &   $-$SD   \\ \hline
                   Muon identification & 0.01 & $-0.01$ \\ 
              Electron identification and smearing & 0.01 & $-0.01$ \\ 
                                     PDF & 0.02 & $-0.01$ \\ 
                                   Top Mass & 0.01 & $-0.01$ \\ 
                                          Triggers & 0.02 & $-0.02$ \\ 
                       Opposite charge requirement & 0.00 & $-0.00$ \\ 
                                  Jet energy scale & 0.01 & $-0.01$ \\ 
                Jet reconstruction and identification & 0.06 & $-0.06$ \\ 
                                    Normalization & 0.02 & $-0.02$ \\ 
                            Monte Carlo statistics & 0.02 & $-0.02$ \\ 
                Instrumental background & 0.00 & $-0.00$ \\ 
                              Background Model for Spin & 0.03 & $-0.04$ \\ 
                                    Luminosity & 0.03 & $-0.03$ \\ 
                                             Other & 0.01 & $-0.01$ \\ 
                 Template statistics for template fits & 0.07 & $-0.07$ \\ 
                                 \hline  
                   Total systematic uncertainty & 0.11 & $-0.11$ \\ \hline
   Statistical uncertainty    & 0.38 & $-0.40$  \\ \hline \hline
 \end{tabular}
\label{tab:template_errors}
\end{center}
\end{table}

While the measured value of $C$ agrees with the predicted value within two
standard deviations, it is also consistent with no spin correlation at
all.  A more powerful technique is required to observe the effect with the
data sample in hand.

\section{Analysis II: Matrix-element-based}
The second measurement makes use of the leading-order matrix element for
$t\bar{t}$ production and decay, using the full event kinematics to
determine the fraction of $t\bar{t}$ events in the sample that the spin
correlation that is expected in the SM~\cite{bib:meas2}.  Matrix
elements have never been used in spin-correlation measurements before, and
their implementation leads to a significant improvement in sensitivity.

On an event by event basis, we can characterize whether the kinematics are
consistent with the SM or with no correlation at all.  This is done by
calculating a probability for consistency of the event with spin
correlation or non-correlation. The probability is given by
\begin{eqnarray}
\label{eq:psgn}  \nonumber
P_{\rm sgn}(x;H) & & \hspace{-0.3cm}= \frac{1}{\sigma_{\rm obs}} \int
~f_{\rm PDF}(q_1)~f_{\rm PDF}(q_2) 
{\rm d}q_1 {\rm d}q_2 \\ 
&\cdot & \frac{(2 \pi)^{4} \left|{\cal M}(y,H)\right|^{2}}
       {q_1q_2 s}
  ~ W(x,y) 
~     {\rm d}\Phi_{6}.
\end{eqnarray}
Here, $\sigma_{\rm obs}$ denotes the leading order cross section including
selection efficiency, $q_1$ and $q_2$ the energy fraction of the incoming
quarks from the proton and antiproton, respectively, $f_{\rm PDF}$ the
parton distribution function, $s$ the center-of-mass energy squared of the
$p\bar{p}$ system and ${\rm d}\Phi_{6}$ the infinitesimal volume element of
the 6-body phase space.  The detector resolution is taken into account
through a transfer function $W(x,y)$ that describes the probability of a
partonic final state $y$ to be measured as $x =
(\tilde{p}_1,\dots,\tilde{p}_n)$, where $\tilde{p}_i$ denotes the measured
four-momenta of the final state particles.  $H$ represents the correlation
hypothesis -- $H=c$ corresponds to SM correlation and $H=u$ corresponds to
no correlation.  The appropriate matrix element is used for each case.  In
contrast to the template measurement, the full event kinematics plus
theoretical models of $t\bar{t}$ production and decay are used, not just
the lepton angles, and by adding this information, the sensitivity of the
measurement is increased.

\begin{figure*}[ht]
\centering
\includegraphics[width=80mm]{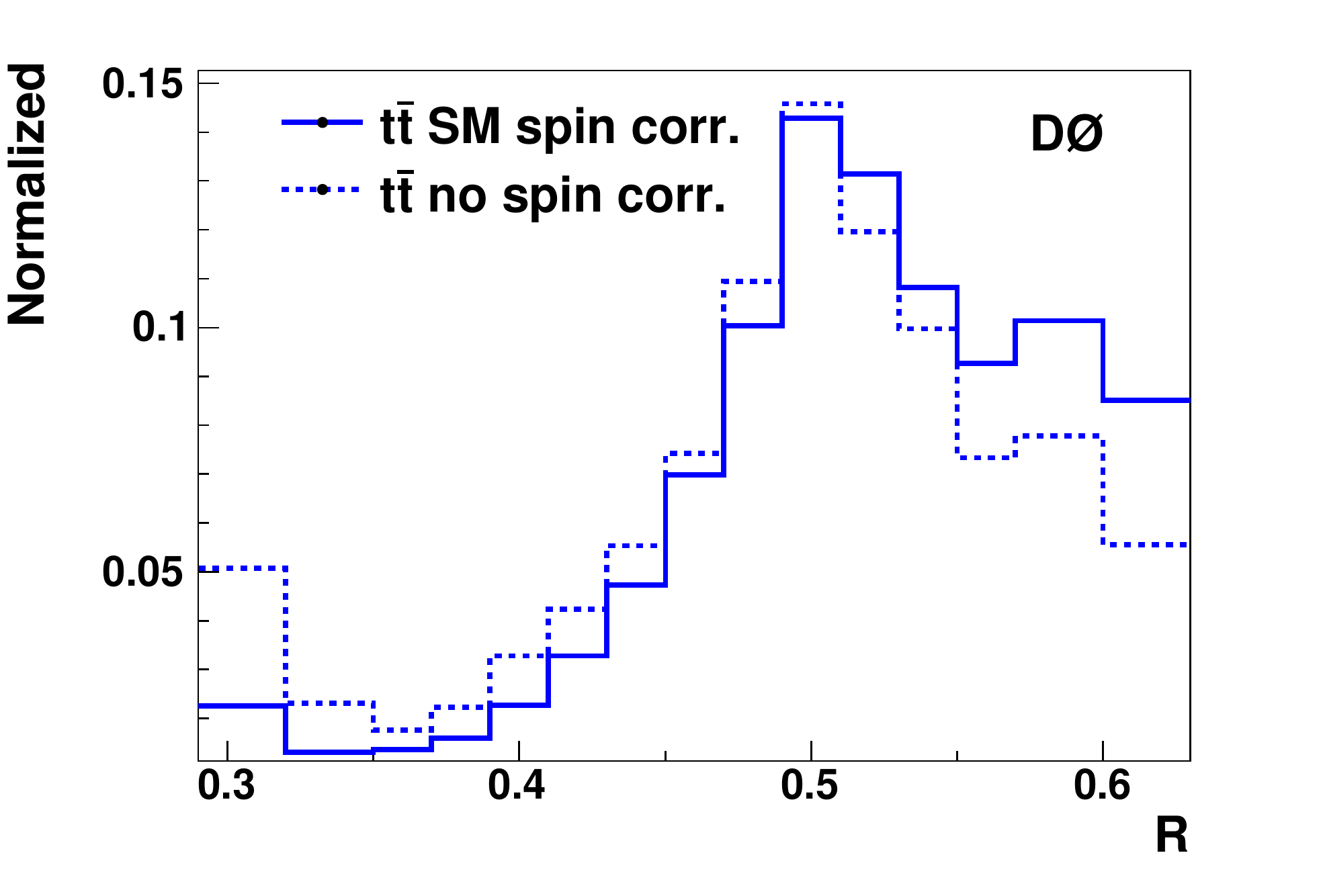}
\includegraphics[width=80mm]{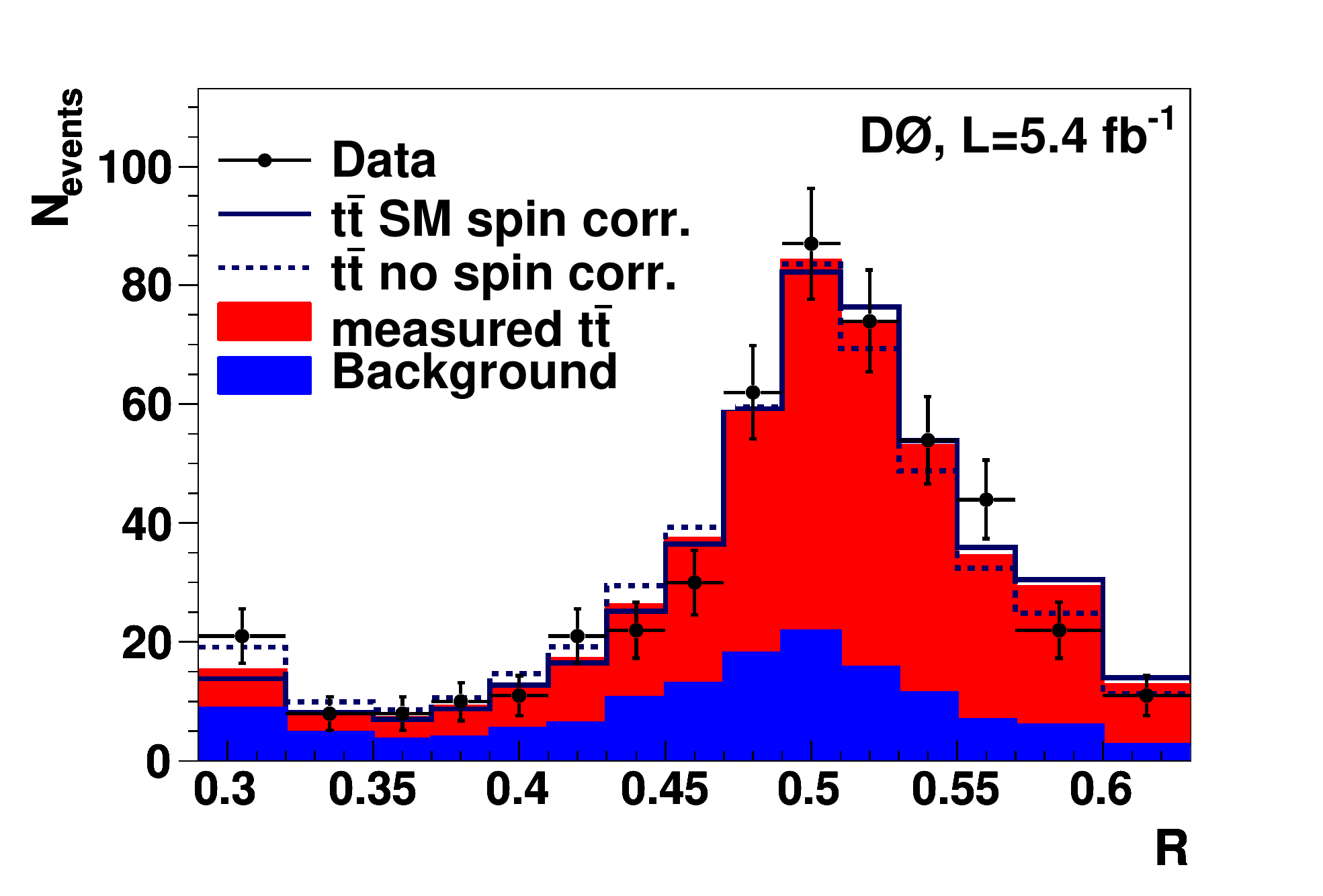}
\caption{Left: Comparison of the discriminant $R$ between SM spin
  correlation $H=c$ and no spin correlation $H=u$ at parton level.  The
  first and last bin include also the contributions from $R<0.29$ and
  $R>0.63$.  Right: The predicted discriminant distribution $R$ for the
  combined dilepton event sample for the fitted $\sigma_{t\bar{t}}$ and
  $f_{\mathrm{meas}}$ compared to the data.  The prediction with spin
  correlation ($f=1$) and without spin correlation ($f=0$) is shown
  including background.} \label{fig:matrix}
\end{figure*}

For each event, we compute
\begin{equation}
R = \frac{P_{\rm sgn} (H=c)}{P_{\rm sgn}(H=u) + P_{\rm sgn} (H=c)}\,,
\label{eq:discr}
\end{equation}
Events more consistent with having SM spin correlation will tend to have
$R$ close to one, while those less consistent will have $R$ closer to zero.
A value of $R \simeq 0.5$ implies that it is is difficult to tell which
hypothesis is more likely.  The left panel of Figure~\ref{fig:matrix} shows
the expected distribution of $R$ for MC@NLO $t\bar{t}$ events generated
with and without spin correlation.  In fact, $R \simeq 0.5$ is quite
common; the correlation is still a small effect.  But there is some
separation between the two distributions.

The right panel of Figure~\ref{fig:matrix} shows the distribution in $R$
observed in the data, along with the distribution expected for the
background events and those for $t\bar{t}$ with and without spin
correlation.  By eye, one can see that the data are more consistent with
the hypothesis of correlation.

\begin{figure*}[ht]
\centering
\includegraphics[width=80mm]{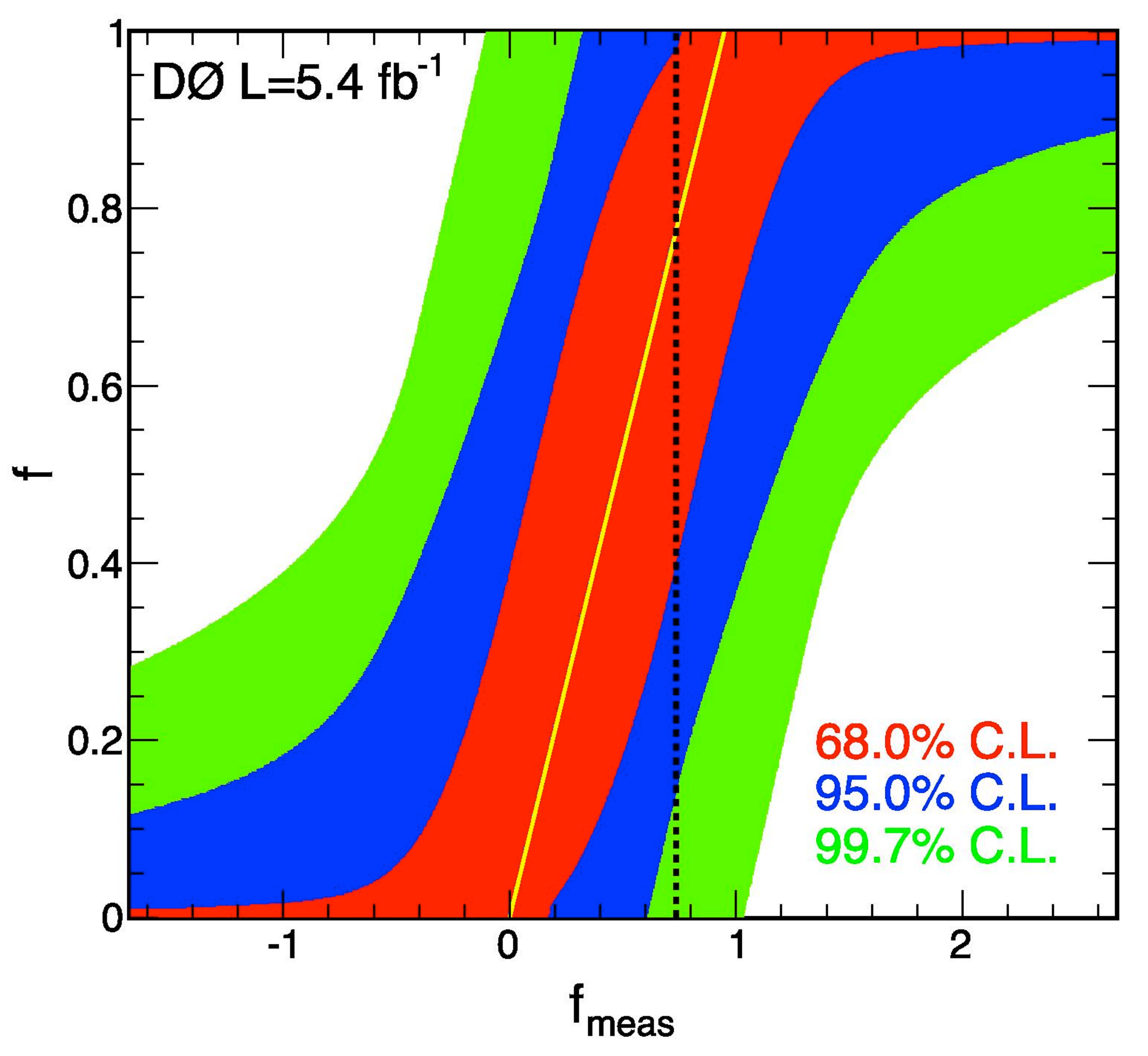}
\caption{The 68.0\% (inner), 95.0\% (central), and 99.7\% (outer)
  C.L. bands of $f$ as a function of $f_{\mathrm{meas}}$ from likelihood
  fits to MC events.  The thin yellow line indicates the most probable
  value of $f$ as a function of $f_{\mathrm{meas}}$, and therefore
  represents the calibration of the method.  The vertical dashed black line
  indicates the measured value $f_{\mathrm{meas}}=0.74$.}
\label{fig:matrixFC}
\end{figure*}

The fraction of events that are consistent with SM spin correlation, $f$,
is obtained from a binned likelihood fit that is very similar to that of
Analysis~I.  This fraction is of course expected to be 100\%.  The
confidence bands are shown in Figure~\ref{fig:matrixFC}.  We find $f =
0.74^{+0.40}_{-0.41}$, which is consistent with the SM.  A value of $f=0$
is excluded at the 97.7\% confidence level.  (From ensemble testing, the
measurement was expected to exclude $f=0$ at the 99.6\% confidence level;
if the correlation exists as expected in the SM, this measurement was
``unlucky'' in finding a value less than 100\%.)  The uncertainties on the
measured value of $f$ are listed in Table~\ref{tab:matrix_errors}; once
again, statistical uncertainties greatly dominate.

\begin{table}[ht]
\begin{center}
\caption{Summary of uncertainties on $f_{\mathrm{meas}}$.}
\begin{tabular}{ccc} \hline \hline
 Source  &        $+$SD    &   $-$SD   \\ \hline
                 Muon identification & 0.01 & -0.01 \\ 
              Electron identification and smearing & 0.02 & -0.02 \\ 
                                   PDF & 0.06 & -0.05 \\ 
                                       $m_t$ & 0.04 & -0.06 \\ 
                                          Triggers & 0.02 & -0.02 \\ 
                       Opposite charge selection & 0.01 & -0.01 \\ 
                                       Jet energy scale & 0.01 & -0.04 \\ 
            Jet reconstruction and identification & 0.02 & -0.06 \\ 
                                     Background normalization & 0.07 & -0.08 \\ 
                            MC statistics & 0.03 & -0.03 \\ 
                    Instrumental background & 0.01 & -0.01 \\ 
                                    Integrated luminosity & 0.04 & -0.04 \\ 
                                             Other & 0.02 & -0.02 \\ 
            MC statistics for template fits & 0.10 & -0.10 \\ 
                                 \hline  
                    Total systematic uncertainty & 0.15 & -0.18 \\ 
\hline  
  Statistical uncertainty &       0.33 &   -0.35  \\ \hline \hline
\end{tabular}
\label{tab:matrix_errors}
\end{center}
\end{table}

\section{Summary}
Quark spin correlation is a phenomenon that can only be seen in $t\bar{t}$
production, thanks to the short top lifetime.  However, it is a subtle
effect that requires large data samples and sophisticated analysis
techniques to observe.  Indeed, the matrix-element technology is perhaps
the most powerful, and most complex, analysis tool that is currently
available for Tevatron data analyses, and it was required here to have the
hope of observing the effects of interest.  Two analyses of $t\bar{t}$
dilepton events at D0 have been performed.  One was a template-based
analysis using full reconstruction of top decays, giving a result within
two standard deviations of the NLO QCD prediction, but also compatible with
the no-correlation hypothesis.  The other one was a matrix-element-based
analysis that gives a result consistent with the SM hypothesis, and
powerful enough to exclude the no-correlation hypothesis for the first time
ever.  Both analyses are statistics limited, with only about half of the
final D0 Run~II data sample analyzed so far.  Thus, there is great
potential for improving the precision of the measurements in the near
future.

\begin{acknowledgments}
I thank the D0 ``spinners'' (Alexander Grohsjean, Tim Head, Yvonne Peters
and Christian Schwanenberger) for their advice while I was preparing this
presentation, and the D0 Collaboration for giving me the opportunity to
present these interesting results.  I also thank the organizers of the DPF
2011 conference for an engaging and enjoyable week.
\end{acknowledgments}

\bigskip % extra skip inserted

\end{document}